\documentclass[letter,traditabstract]{aa}
\usepackage{txfonts}
\usepackage{graphicx}
\usepackage{float}
\usepackage{natbib}
\usepackage{hyperref}
\usepackage[english]{babel}
\usepackage{color}
\usepackage{hyperref}
\usepackage[flushleft]{threeparttable}

\title{The influence of wavelength, flux, and lensing selection effects on the redshift distribution of dusty, star-forming galaxies}

\author{Matthieu~B{\'e}thermin\inst{1} \and Carlos De Breuck\inst{1} \and Mark Sargent\inst{2} \and Emanuele Daddi\inst{3}}
\institute{European Southern Observatory, Karl-Schwarzschild-Str. 2, 85748 Garching, Germany \email{mbetherm@eso.org} \and  Astronomy Centre, Department of Physics and Astronomy, University of Sussex, Brighton, BN1 9QH, UK \and Laboratoire AIM-Paris-Saclay, CEA/DSM/Irfu - CNRS - Universit\'e Paris Diderot, CEA-Saclay, pt courrier 131, F-91191 Gif-sur-Yvette, France}

\date{Received ??? / Accepted ???}

\abstract{We interpret the large variety of redshift distributions of galaxies found by far-infrared and (sub-)millimeter deep surveys depending on their depth and wavelength using the Béthermin et al. (2012) phenomenological model of galaxy evolution. This model reproduces without any new parameter tuning the observed redshift distributions from 100\,$\mu$m to 1.4\,mm, and especially the increase of the median redshift with survey wavelength. This median redshift varies also significantly with the depth of the surveys, and deeper surveys do necessarily not probe higher redshifts. Paradoxically, at fixed wavelength and flux limit, the lensed sources are not always at higher redshift. We found that the higher redshift of 1.4\,mm-selected south pole telescope (SPT) sources compared to other SMG surveys is not only caused by the lensing selection, but also by the longer wavelength. This SPT sample is expected to be dominated by a population of lensed main-sequence galaxies and a minor contribution ($\sim$10\%) of unlensed extreme starbursts.}


\titlerunning{The influence of selection effects of redshift distribution of SMGs}

\authorrunning{B\'ethermin et al.}

\begin{document}

\maketitle

\section{Introduction}


The determination of the star formation history in the Universe is a key challenge for understanding the evolution of galaxies \citep{Madau2014}. Two decades ago, the first deep submillimeter surveys (850\,$\mu$m) revealed a population of dusty, strongly star-forming galaxies \citep[e.g.,][]{Smail1997}, which were missed by optical surveys. This showed the importance of submillimeter observations to draw a complete picture of the star formation activity in the high-redshift Universe. \citet{Chapman2005} found that the median redshift of these submillimeter galaxies (SMGs) is 2.3. However, the identification of the optical counterparts was non trivial because of the large beam of single-dish submillimeter telescopes ($\sim$15-20"). Their precise position was determined by radio interferometry and their redshift measured using optical spectroscopy. This method was thus potentially biased against the highest redshift objects and galaxies in the 1.4$<$z$<$2 redshift desert.

A decade later, \citet{Vieira2010,Vieira2013} identified a population of strongly-lensed dusty star-forming galaxies (DSFGs) in the 1.4\,mm South Pole Telescope (SPT) survey. Using ALMA to derive directly spectroscopic redshifts by targeting CO-transitions at millimeter wavelengths, \citet{Weiss2013} found that their median redshift is 3.5. \citet{Simpson2014} used millimeter interferometry with a subarcsec resolution to directly confirm the optical counterparts of SMGs and measured a median photometric redshift similar to \citet{Chapman2005}. Is this difference of median redshift caused by the different wavelengths or the lensing effect? In this letter, we discuss these selection effects using the \citet[][hereafter B12]{Bethermin2012c} model, that reproduces the redshift distribution of sources detected at all wavelengths from 100\,$\mu$m to 2\,mm (Sect.\,\ref{sect:model}). Using a simplified version of this model, we explain why sources selected at longer wavelengths have a higher median redshift (Sect.\,\ref{sect:hlhz}). We then show how the redshift distribution of DSFGs is affected by the flux and lensing selection (Sect.\,\ref{sect:seleff}). In Sect.\,\ref{sect:spt}, we discuss in more detail the SPT-selected population of galaxies in the light of our findings of the two preceding sections. We define L$_{\rm IR}$ as the bolometric energy emitted by a galaxy between 8 and 1000\,$\mu$m.

\begin{figure}
\centering 
\includegraphics[scale=1.0]{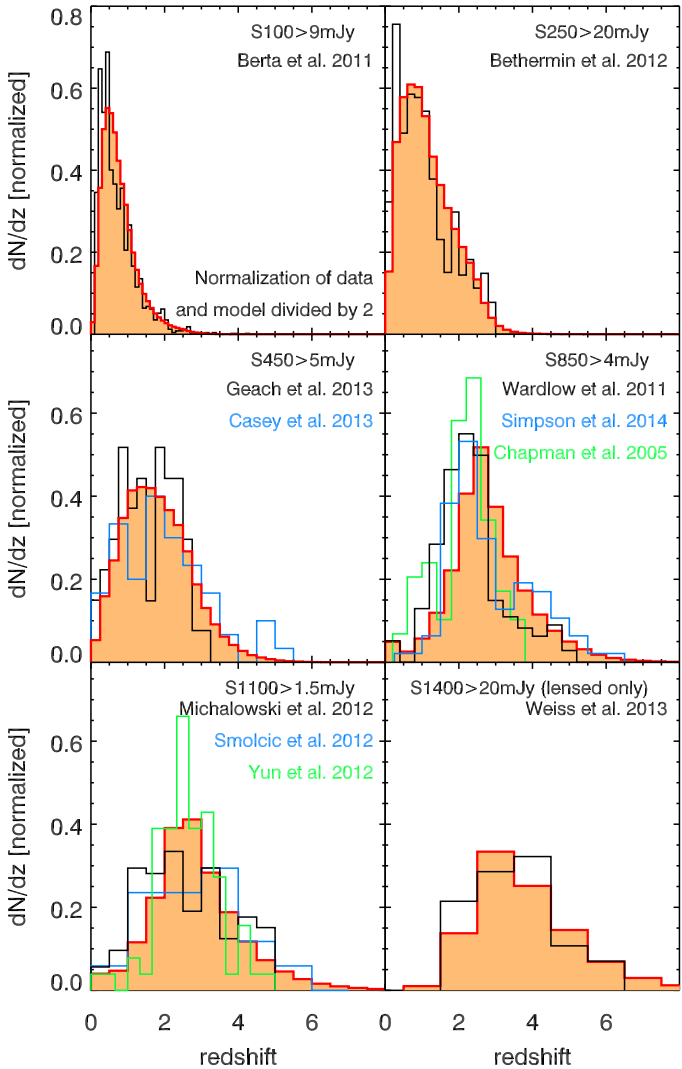}
\caption{\label{fig:comp} Comparison between the prediction of our model (orange filled histogram) and the observations (see Table\,\ref{tab:nz}). The flux cuts indicated to compute the redshift distributions are indicated on the figures. The distributions are normalized to have $\int (dN/dz) \, dz = 1$ (expect the 100\,$\mu$m normalized at 0.5 for clarity).}
\end{figure}

\section{Model and comparison with observations}

\label{sect:model}

The B12 is model is based on the observed evolution of the stellar mass (M$_\star$) function and the main-sequence of star-forming galaxies, i.e., a tight SFR-M$_\star$ correlation. At fixed mass, the star formation rate (SFR) increases rapidly but smoothly with redshift. This model also includes a population of starbursts with a strong excess of sSFR (=SFR/M$_\star$) that contributes 15\% of the star formation density at z$>$1 \citep{Sargent2012}. We used two different SEDs based on \citet{Magdis2012b} for main-sequence and starburst galaxies. The dust of the main-sequence templates is warmer at higher redshifts. The contribution of strongly-lensed galaxies is computed using the model of \citet{Hezaveh2011}. The source size can affect the lensing amplification, but an extreme size evolution by a factor of 12.5 between $z$=2.5 and $z$=6 would be required to match the 850\,$\mu$m and SPT redshift distributions \citep{Weiss2013}. Recent observations \citep{Ikarashi2014,Smolcic2014,Simpson2015} do not suggest size evolution. We thus do not adopt any size evolution in our model. In this letter we focus on the 100-2000\,$\mu$m range, since at shorter and larger wavelengths the AGN have a non-negligible contribution through their torus and/or synchrotron emissions \citep{Drouart2014}.

We test the validity of our model by comparing it with a large compilation of redshift distributions listed in Table\,\ref{tab:nz}. Overall the model reproduces well the observed redshift distributions from 100\,$\mu$m to 1.4\,mm (see Fig.\,\ref{fig:comp}). At 2\,mm (not shown in Fig.\,\ref{fig:comp}), only 5 of the of the 12 sources found by \citet{Staguhn2014} have known redshifts. Our model predicts a median redshift of 2.9 assuming of flux cut of 0.24\,mJy\footnote{The flux density of their faintest source after deboosting.}, while they found 2.9$\pm$0.4. Note that the good agreement between our model and the data is not the result of fine tuning the ingredients of our model. We can nevertheless observe a small tension at 850\,$\mu$m, where our distribution peaks at a redshift too high by $\Delta$z$\sim$0.2. At the flux density cuts used for all panels in Fig\,\ref{fig:comp}, our model predicts that we mainly select main-sequence galaxies. The smooth evolution of the main-sequence \citep{Schreiber2014} and the mass function \citep{Ilbert2013}, as well as the volume effects are thus sufficient to explain the redshift distribution from 100\,$\mu$m to 1.4\,mm. \citet{Zavala2014} extrapolated successfully the redshift distributions from 850\,$\mu$m up to 1.4\,mm, but did not manage to reproduce the 450\,$\mu$m. They claimed another galaxy population is necessary. This could be caused by their assumption of a flux-invariant redshift distribution, which is not present in our more refined model (see Sect.\,\ref{sect:seleff}). 

\begin{table}
\caption{\label{tab:nz} Summary of data used in Fig.\,\ref{fig:comp}}
\begin{tabular}{llllr}
\hline
\hline
Reference & N & $\lambda_{\rm obs}$ & S$_{\rm lim}$ & Method \\
 & & $\mu$m & mJy & \\
\hline
\citet{Berta2011} & 5360 &100 & 9 & a, b\\
\citet{Bethermin2012b} & 2517 & 250 & 20 & a, b\\
\citet{Geach2013} & 60 & 450 & 5 & b \\
\citet{Casey2013} & 78 &450 & 13 & b \\
\citet{Wardlow2011} & 72 & 850& 4 & b\\
\citet{Simpson2014} & 77 & 850 & 4 & b, e\\
\citet{Chapman2005} & 73 & 850 & 3 & c\\
\citet{Smolcic2012} & 28 & 1100& 1.4 & b, e\\
\citet{Michalowski2012} & 95 & 1100 & 1 & b \\
\citet{Yun2012} & 27 &1100 & 2 & b\\
\citet{Weiss2013} & 23 & 1400 & 20 & d, f\\
\citet{Staguhn2014} & 5 & 2000 & 0.24 & b, c\\
\hline
\end{tabular}
\tablefoot{a) extraction of the sources based on PSF-fitting codes using short-wavelength priors; b) photometric redshifts; c) optical/near-infrared spectroscopic redshift after radio identification; d) millimeter spectroscopic redshift; e) identification of the optical/near-infrared counterparts using high-resolution (sub-)millimeter data.}
\end{table}

\section{Why does selection at longer wavelengths select higher-redshift galaxies?}

\label{sect:hlhz}

\begin{figure*}
\centering 
\includegraphics[scale=1.0]{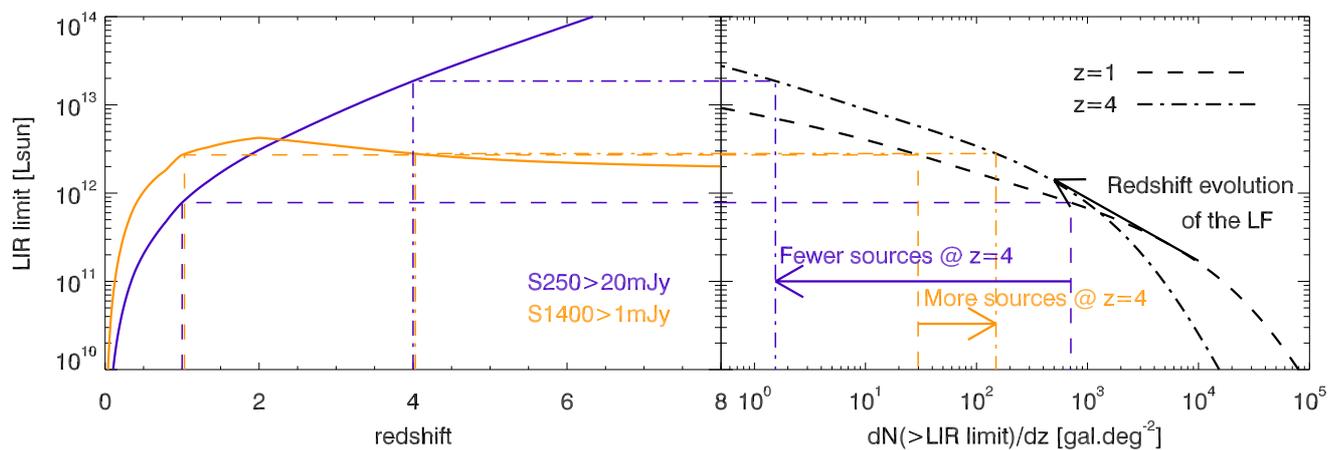}
\caption{\label{fig:simplified} Simplified diagram illustrating the impact of the wavelength on the redshift distribution. The left panel represents the detection limit in bolometric infrared luminosity ($L_{\rm IR}$) as a function of redshift for a S$_{250}>$20\,mJy (purple) and S$_{1400}>$1\,mJy (gold). These limits are computed using the \citet{Magdis2012b} main-sequence SED library. We neglected the scatter and the starburst populations to simplify this diagram. The right panel represents the number density of objects per redshift interval (computed from the luminosity function and the geometry of the Universe) as a function of the $L_{\rm IR}$ cut (x and y axis are inverted) at z=1 (dashed line) and z=4 (dot-dash line). These curves are computed using our model. The colored arrows represent the ratio between the number density of z=1 and z=4 sources. The black arrow highlights the increase of the luminosity and the decrease of the density of the knee of the LF between z=1 and z=4.}
\end{figure*}

The shift of the peak of the redshift distribution towards higher redshift when wavelength increases (Fig.\,\ref{fig:comp}) can be easily explained using a simplified version of our model. In this version, we neglect the effect of the strong lensing and assume that all galaxies at a given redshift have the average main-sequence SED provided by the \citet{Magdis2012b} template library. The left panel of Fig.\,\ref{fig:simplified} shows how the L$_{\rm IR}$ limit at which sources can be detected evolves with redshift for the flux density cuts used in Fig.\,\ref{fig:comp}. At z$>$1, the curve is almost flat at 1.4\,mm, but increases quickly with increasing redshift at 250$\mu$m. This effect is caused by the shift of the peak of the dust emission ($\sim$100\,$\mu$m rest-frame) toward the observed millimeter wavelengths when the redshift increases \citep[e.g.,][]{Blain2002}.

The redshift distribution can then be deduced from the luminosity function (LF). The right panel of Fig.\,\ref{fig:simplified} shows the cumulative luminosity functions at z=1 and z=4 computed from the B12 model. These functions are renormalized to directly provide the number of sources per square degree and redshift interval above a certain L$_{\rm IR}$ cut. The x-axis and y-axis switched with respect to the usual representation to better illustrate the connection with the left panel. At 250\,$\mu$m, the L$_{\rm IR}$ limit of surveys is 24 times larger at z=4 than at z=1. Consequently, despite a slightly higher density of luminous objects at z=4, we detect 450 times fewer objects at z=4, because of the steep slope of the luminosity function above the knee. At 1.4\,mm, the L$_{\rm IR}$ limits are similar at both redshift, but we detect 5 times more objects at z=4, because of the evolution of the LF.

\section{Impact of flux and lensing on the redshift distribution}

\label{sect:seleff}

\begin{figure}
\centering 
\includegraphics[scale=0.93]{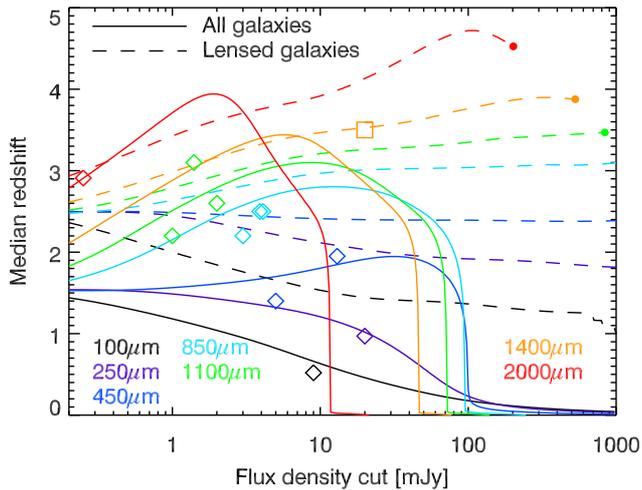}
\caption{\label{fig:zmed} Median redshift of dusty galaxies as a function of the flux cut at various wavelengths (see color coding in the plot). The solid lines correspond to full samples and dashed lines to strongly-lensed samples. The filled dot indicates the limit of one source per 2$\pi$\,sr (about half of the sky is sufficiently clean for extragalactic surveys). The observations listed in Table\,\ref{tab:nz} are symbolized by diamonds (full samples) and a square (SPT lensed sample).}
\end{figure}

We also used the standard version of our model (including strong lensing, different templates for main-sequence and starburst galaxies, and temperature dispersion) to predict more precisely how the median redshift of galaxies selected by photometric surveys changes depending on the wavelength and the flux density cut (Fig.\,\ref{fig:zmed}). 

At $\lambda \ge$450\,$\mu$m, we found an abrupt decrease of the median redshift around 100\,mJy (10\,mJy at 2\,mm). This sharp transition is caused by the presence of two bumps in the redshift distributions: local galaxies around the knee of the luminosity function and luminous galaxies z$\sim$2-3. There are fewer sources at z$\sim$0.5-1 than at z$\sim$0.1, because the L$_{\rm IR}$ limit is well above the knee of the luminosity function. There are also more sources at z$\sim$2-3 than at z$\sim$0.5-1, because the volume probed is larger and the number of luminous galaxies per unit of volume higher (see, e.g., \citealt{Bethermin2011}). At high fluxes, the number of nearby objects decreases following an Euclidian trend in S$^{-5/2}$ (see, e.g. \citealt{Planck_eucl}). At higher redshift, we are probing objects above the knee of the luminosity function, where the slope is exponential. Consequently, when we observe at very high flux, the nearby population always dominates the redshift distribution (e.g. the S$_{500}>100$\,mJy sources), while the lensed population has a median redshift of $\sim$2.4 \citep{Negrello2010, Gonzalez2012}.

We now consider the redshift distribution of only the strongly-lensed galaxies (dashed line in Fig.\,\ref{fig:zmed}). At $\lambda$$<$1.1\,mm, the lensed galaxies are always at higher redshift than the full population. This is expected, since the probability of lensing increases with redshift because of the larger probability to find a massive galaxy on the line of sight (see, e.g., \citealt{Hezaveh2011}). However, at $\lambda$$>$1.1\,mm, the lensed objects are at a lower median redshift in some specific flux range (2-7\,mJy at 1.4\,mm). This can be explained by the lower intrinsic fluxes ($<$1\,mJy) of the lensed sources. The redshift evolution of the luminosity function and L$_{\rm IR}$ limit (Fig.\,\ref{fig:simplified}) puts these faint sources at lower median redshift than the $>$1\,mJy sources which dominate the unlensed population. In the domain of a few mJy, this effect can thus compensate the lensing effect, which biases the redshift distributions toward higher redshift.


These results suggests that the best strategy to build samples of high-redshift galaxies is to perform surveys at the largest possible wavelength. However, all objects below z=8 have similar 1.4\,mm/2\,mm colors, because they are all observed in the Rayleigh-Jeans regime. The flux at 2\,mm is 3 times lower, while the sensitivity of instruments is similar for the two wavelengths. Consequently, the expected number of detections for the same integration time is significantly lower. The higher median redshift at 2\,mm is mainly caused by the lack of L$_{\rm IR}$ sensitivity at z$<$2 rather than a better efficiency to detect very high redshift sources. A compromise has thus to be found between the large wavelength and the efficiency of the survey, defined here as the number of detections per hour. Furthermore, the risk of contamination by free-free emission increases and source identification at longer wavelengths is more difficult due to the larger beam. At $\lambda > 500$\,$\mu$m, the median redshift decreases, when we go deeper than 1 mJy. Deeper surveys do not automatically imply higher redshifts.

\section{A closer look at the SPT galaxy population}

\label{sect:spt}

\begin{figure}
\centering 
\includegraphics[scale=0.93]{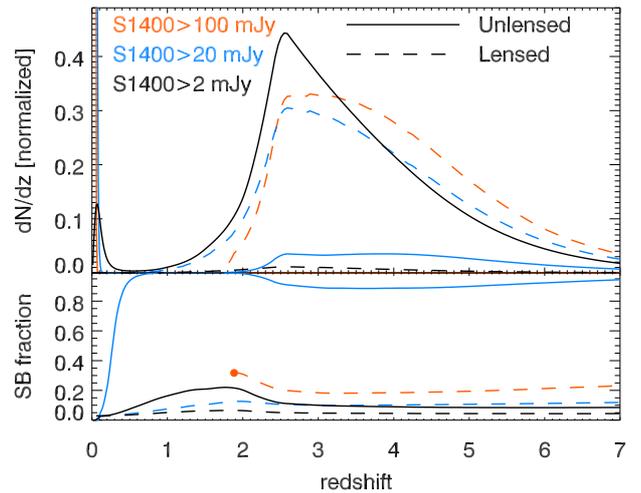}
\caption{\label{fig:nz_spt} Upper panel: redshift distribution from the B12 model of dusty galaxies selected at 1.4\,mm for various flux cuts (see color coding in the plot). The unlensed galaxies are represented by solid lines and the contribution of lensed populations by a dashed line. For each flux cut, we normalized the sum of the lensed and unlensed distributions considering only z$>$0.1 sources to allow an easier comparison of high redshift distributions. Lower panel: fraction of starbursts. The curves are not plotted, where dN/dz$<$1\,sr$^{-1}$.}
\end{figure}

The SPT survey found a large population of lensed galaxies using a selection at 1.4\,mm \citep{Vieira2010}. \citet{Weiss2013} found that their median redshift is the highest measured to date in a simple photometric selection. We will discuss their characteristics in this section. Fig.\,\ref{fig:nz_spt} shows the redshift distribution predicted by the B12 model of sources selected at 1.4\,mm with the 20\,mJy density flux cut of \citet{Weiss2013}, but also a lower (1\,mJy) and a higher (100\,mJy) one to illustrate how this influences the nature of the selected sources. The contribution of z$<0.1$ is significant for 20\,mJy and 100\,mJy (32\% and 70\%, respectively), but negligible for 2 mJy (0.9\%). Since we are more interested in the high-z population, we normalized the redshift distribution in Fig.\,\ref{fig:nz_spt} ignoring the objects at z$<$0.1.

The nature of the high-redshift sources also depends strongly on the flux cut. For the 20\,mJy cut, 87\% of sources are lensed and 90\% of them are classified by the model as "main-sequence". Using a mock catalog based on our model, we estimated that their average sSFR is 1.7 times higher than the center of the main-sequence, while in our model, we define starbursts as lying $>4 \times$ above. Because of their higher SFR at fixed stellar mass, the galaxies slightly above ($\sim$1$\sigma$) the center of the main sequence are easier to detect than objects slightly below. The unlensed sources are extreme HyLIRG (L$_{\rm IR}>$10$^{13}$\,L$_\odot$) starbursts with an excess compared to the main sequence larger than a factor of 4. For 100\,mJy, all the high-redshift sources are lensed. This is expected, because the minimal intrinsic luminosity necessary to detect a source without lensing, assuming our coldest template, is 10$^{14}$\,L$_\sun$, which is unphysical. For a cut of 2\,mJy corresponding to the expected sensitivity of SPT-3g \citep{Benson2014}, we typically expect to detect unlensed ULIRGs ($\langle$L$_{\rm IR} \rangle$ = 7$\times$10$^{12}$\,L$_\odot$) at z$\sim$3.2. At this redshift, this corresponds to the knee of the luminosity function. The slope of the counts is thus shallower and the contribution of lensed sources is small (2.9\%).
 
Our model predicts that SPT sources lie only slightly above and still well within the scatter of the main sequence. The high SFR of these objects is expected to be caused by large gas reservoirs rather than a merger-induced, boosted star-formation efficiency \citep{Sargent2014,Bethermin2015}.\\ 

\section{Conclusion}

The B12 model allows us to understand how the observed redshift distributions of DSFGs depend on how they are selected. Our main findings are:
\begin{itemize}
\item The B12 model successfully reproduces the redshift distributions from 100\,$\mu$m to 1.4\,mm without any additional parameter tuning. 
\item When we select sources at longer wavelength, the median redshift of the sources increases. This effect can be easily explained considering the L$_{\rm IR}$ limit versus redshift of the surveys and the evolution of the infrared LF.
\item The median redshift of the DSFGs also varies with survey depth. The deeper surveys in the (sub-)millimeter surprisingly probe lower redshifts. At $\lambda >$1.4\,mm in specific flux intervals, the lensed objects can also be at higher redshifts than unlensed sources.
\item The DSFGs selected by SPT are mainly strongly magnified, main-sequence galaxies, but 10\% of these sources are predicted to be unlensed HyLlRGs.
\end{itemize}  

\begin{acknowledgements}
We acknowledge ERC-StG UPGAL 240039 and ANR-08-JCJC-0008 and Manuel Aravena for his comments.
\end{acknowledgements}


\bibliographystyle{aa}

\bibliography{biblio}

\end{document}